\documentclass[10pt, conference]{IEEEtran}
\IEEEoverridecommandlockouts

\usepackage{subfigure}

\usepackage[utf8]{inputenc}
\usepackage{amsmath,amssymb}

\usepackage{graphicx}
\usepackage{subfigure}
\usepackage{color}
\usepackage{array}

\usepackage{algorithm}
\usepackage{algorithmic}
\usepackage{url}
\usepackage{color}

\begin{document}

\title{Mobile Computing in Digital Ecosystems:\\ Design Issues and Challenges\footnotemark}

\author{\IEEEauthorblockN{Gabriele D'Angelo, Stefano Ferretti, Vittorio Ghini, Fabio Panzieri}
\IEEEauthorblockA{Department of Computer Science, University of Bologna\\
Bologna, Italy\\
\{gdangelo, sferrett, ghini, panzieri\}@cs.unibo.it}
}


\maketitle

\footnotetext{The publisher version of this paper is available at \url{http://dx.doi.org/10.1109/IWCMC.2011.5982863}. \textbf{{\color{red}Please cite as: Gabriele D'Angelo, Stefano Ferretti, Vittorio Ghini, Fabio Panzieri. Mobile Computing in Digital Ecosystems: Design Issues and Challenges. Proceedings of the 7th International Wireless Communications and Mobile Computing conference (IWCMC-2011), Emergency Management: Communication and Computing Platforms Workshop. ISBN 978-1-4577-9538-2.}}}

\begin{abstract}
In this paper we argue that the set of wireless, mobile devices
(e.g., portable telephones, tablet PCs, GPS navigators, media players)
commonly used by human users enables the construction of what we term a ``digital ecosystem'',
i.e., an ecosystem constructed out of so-called ``digital organisms'' (see below),
that can foster the development of novel distributed services. 
In this context,
a human user equipped with his/her own mobile devices,
can be though of as a ``digital organism'' (DO), 
a subsystem characterized by a set of peculiar features and resources it can offer to the rest of the ecosystem for use from its peer DOs. 
The internal organization of the DO must address issues of management of 
its own resources, 
including power consumption. 
Inside the DO and among DOs,
peer-to-peer interaction mechanisms can be conveniently deployed to favor resource sharing and data dissemination. 
Throughout this paper, 
we show that most of the solutions and technologies needed to construct a digital ecosystem are already available. 
What is still missing is a framework 
(i.e., mechanisms, protocols, services) 
that can support effectively the integration and cooperation of these technologies.
In addition, 
in the following we show that that framework can be implemented as a middleware subsystem that enables novel and ubiquitous forms of computation and communication.
Finally, 
in order to illustrate the effectiveness of our approach,
we introduce some experimental results we have obtained from preliminary implementations of 
(parts of) 
that subsystem.
computation and communication.
\end{abstract}


\section{Introduction}

The proliferation of mobile, 
heterogeneous devices gives rise to new scenarios that foster the cooperation among individuals 
through the pervasive and ubiquitous use of these devices.
However,
owing to the their mass diffusion,  
new and effective solutions for the management and organization of these devices are required in order to guarantee and optimize ``always on'' services. 
A key feature of these new scenarios is that users may adopt very heterogeneous devices. 
Specifically,
a user may use, 
concurrently,
several mobile devices, such as a portable phone or a tablet PC,
each of which can be characterized by a specific technical profile
(i.e., specific hardware, computational and communication capabilities)
and embedded in a large complex system.\\

Thus,
within these scenarios,
each mobile user can be thought of as a ``digital organism'' (DO) which may interact with peer DOs;
a community of interacting DOs can be thought of as a ``digital ecosystem''.
Each DO may have its own objectives and its own forms of interaction within the ecosystem in which it is embedded.
In addition,
it contributes to maintaining that ecosystem by providing it with its own unused resources. 
Literature on Peer-to-Peer (P2P) systems shows that optimization mechanisms for the sharing of resources 
(intended as data, information, communication, computing)
among cooperating entities can enhance the development of ubiquitous services, 
as long as they attain coordination among those entities. 
Therefore,
a good digital ecosystem should provide the community of DOs it embeds with an environment that facilitates their (self-)organization and interaction. 
Hence,
in principle,
possible heterogeneities among the DOs populating the same ecosystem are not regarded as a problem; 
rather, they provide an opportunity to promote cooperation.
In practice,
these heterogeneities may require that adaptation policies be deployed by the ecosystem itself,
transparently to the DOs it hosts,
in order to enable cooperation among them.\\

Organizing a solid digital ecosystem, 
fully aware of all mobility issues, 
requires that a number of technological limitations be overcome. 
DOs should be made capable of using rationally their resources. 
Most often, 
different devices within the same DO do not interact with each other or, 
if they do, 
interaction may occur only through manual 
(and usually complicated) 
configuration operations carried out by the human user. 
In order to enable DOs to use and share the resources available in their home ecosystem, 
they must be endowed with adaptive mechanisms 
(fully transparent to their human users) 
that can optimize and coordinate the use of those resources. 
In summary,
firstly the different devices in a DO should be able to communicate with each other via short-range wireless technologies. 
Secondly, 
information sent outside a DO should pass through a special ``gateway'' that will select the output interface according to such criteria as ``battery level'', 
for example,
or ``best available communication network''
(e.g., the gateway may give access to a cellular phone interface if the ``best'' available network is a cellular network, 
or a laptop or PDA Wi-Fi interface, if the ``best'' available network is an available Wi-Fi network; note that ``best'' here may mean ``with lowest latency'', 
or ``cheaper''
depending on specific human user QoS requirements). 
Thirdly, 
and finally, 
computation should be performed by the device with the highest computational capability.\\


Once individuals are organized as entities able to optimally move and act within the digital ecosystem, 
they must be enabled to interact with the external environment in the best possible way. 
From this perspective, how to opportunistically exploit all available technological solutions is a key issue. In order to enjoy ``always on'' services, users should be able to move transparently from one network to another without losing their active connections.

After individuals have been endowed with the needed technologies for adaptive living in the digital ecosystem, it is natural to envision several forms of cooperation among users, in order to enhance and globally optimize the overall resource use for given user-specific needs. In particular, users can freely organize themselves into ad-hoc networks (MANETs) and share resources. For example, a user might decide to share his laptop's Wi-Fi network interface with a neighbor user; the devices in this latter user accessing that interface will communicate with it via Bluetooth or infrared channels. This technique can be extended so as to allow the interaction between devices for handling computations on remote hosts.
The substantial difference between the digital ecosystem model of interaction to be developed and the traditional ones is that each DO in an ecosystem is to be regarded as a possible computational and communication resource, not simply as a node in a network that has to forward messages towards their final destination (a feature which will nevertheless be guaranteed in the digital ecosystem).\\

In this paper, we discuss issues of design of a digital ecosystem to be deployed in pervasive environments. 
The ecosystem can be developed and optimized based on a new paradigm we propose; this paradigm is in turn based on self-organizing, opportunistic networking, and the P2P communication model.
The approach we propose can be summarized in the following four steps:
\begin{enumerate}
 \item Resources available at a given mobile user must be integrated using autoconfiguration strategies. In particular, the Personal Area Network (PAN) should be configured so that all the data generated/collected at a given device can be made available to other devices of the same user, ready to be used for any type of computation. The configuration should be accomplished taking into consideration power consumption issues. Such a dynamic integration converts a set of owned devices into a real DO that can exploit all the technological features available to the user. Furthermore, in the dynamic configuration strategies, it will be necessary to consider user level policies such as the cost issues related to service usage (e.g. mobile and broadband connectivity).
 \item The DO needs to be equipped with a software module, running on a given device of the DO, which acts as a gateway that exploits multi-criteria, adaptive decision schemes to understand which is the best network interface to use to send/receive information to/from the outside world.
 \item Smart P2P schemes must be employed among DOs to share data/resources. These schemes should take into consideration: i) social aspects (e.g.~users might want to share resources only with their friends); ii) trust, security and privacy issues; iii) smart discovery strategies to identify neighbors owning resource/data of interest; iv) tit-for-tat schemes \cite{Menasche:2010}.
 \item While interactions among digital organisms are purely local, such a local organization should reflect a wider view of the global overlay, so as to structure the network based on some desired topology \cite{gridpeer}. Indeed, the structure of the network is of paramount importance to guarantee effective data dissemination \cite{disio}.
\end{enumerate}

This paper is organized as follows. In the next Section we introduce the background related to the topics addressed in this paper. Section \ref{sec:des} discusses our objectives. Section \ref{sec:archi} introduces the principal issues to be addressed in the design of an effective digital ecosystem and illustrate some performance results we have obtained from a partial implementation of an ecosystem. Finally, Section \ref{sec:conc} provides some concluding remarks.

\section{Background}
\label{sec:back}

The massive and pervasive use of personal computers as well as other devices such as smart cellphones, PDAs, new generation household appliances and the like, combined with the widespread availability of Internet connections on one side and wireless networks on the other, raise new important research issues, commonly grouped together under the caption of ``ubiquitous computing''. A large community of researchers is investigating these issues, and a host of technologies for ubiquitous computing has indeed been developed (for the sake of conciseness, we shall not examine them in this paper). 
For instance, both the project ``seamless computing'' \cite{chalmers2003a} proposed by Microsoft in 2003, and Jini, which is a part of the Java technology developed by Sun Microsystems, have addressed issues of ubiquitous computing. Both these initiatives stimulated an initial strong interest. However, they were rapidly forgotten by the research and software developers communities as they did not introduce any solution or product worth of mention. 
In the context of large scale computing architectures, metacomputing \cite{foster} has been another attempt to propose a new paradigm for the organization of large computer networks. Although metacomputing has achieved rather promising results, a main problem with it, which is still open, is the lack of integration between the global infrastructure and the users' devices.

\subsection{Interaction And Sharing Mechanisms}

Several interaction mechanisms exist for the sharing and exchange of resources and information. Seti@Home is probably one of the first system that implements distributed computing on different hosts. In addition, several techniques exist today for the distributed resource utilization; these techniques are usually referred to as ``cloud computing'' or ``ad-hoc data processing'' \cite{Hayes:2008}.

As to the resources discovery, several P2P alternatives exist such as publish-subscribe systems, DHT (Distributed Hash Table) based solutions, application level multicast \cite{Eugster:2003}.

As to data dissemination, overlay nets are useful to deliver messages among a set of nodes in a distributed system. However, when one tries to adopt such an approach in a mobile scenario, it must create a mobile ad-hoc network (MANET). In fact, a MANET lets nodes communicate among themselves even when no access points are available. The problem is that when an overlay network among mobile nodes is to be created, additional requirements emerge, such as those related to the limited computational and communication capabilities of mobile nodes, their battery capacities, the fact that nodes may join/leave the network dynamically. It is well-known that the task of building a P2P network over a MANET is not an easy task \cite{GruberSK06}. Indeed, the proximity of nodes often suggests to set them as neighbors in the communication overlay, while, from an application point of view, they do not actually need to communicate and exchange messages. Thus, in order to optimize the communication among nodes, it is worth 
considering both the topology of nodes in real space, as well as the interaction requirements among these nodes at the application level.

\subsection{Use of Multiple Networks}


An important feature to be provided is the ability to allow a mobile node, having multiple wireless network interfaces, to change network points of attachment (handover) without disrupting existing connections. An example of solution to support such a host mobility is Mobile IPv6. This solution, implemented at the network layer, requires support from both network infrastructure and end-nodes. In contrast, Proxy Mobile IPv6 relies on network infrastructure support only. The Host Identity Protocol requires support only from end-nodes. The Terminal Mobility Support Protocol ensures IP address transparency by resorting to external SIP (Session Initiation Protocol) proxies. MMUSE works at the application layer and supports vertical handovers by using SIP proxies, placed at the edge of the access networks \cite{Bonola:2009}. In \cite{Tsukamoto:2008} a scheme is proposed that uses cross-layer information to accomplish seamless handovers. A similar, more general solution is the IEEE 802.21 standard which defines a 
framework (Media Independent Handover services) for the exchange of information among different layers, regardless of the low-level technologies.\\

Research works on ad-hoc networks investigate a wide set of issues, and in particular, radio spectrum management, hierarchical organization, location-based routing, nodes addressing and energy consumption. In this scenario, cognitive radio technology allows a secondary user to share the wireless channel with the primary user in an opportunistic manner. Other proposals focus on the decomposition of ad-hoc networks into small overlapping clusters instead; these are exploited to build a tree-like network, with the aim of prolonging the network lifetime through a more efficient energy utilization.\\

Finally, the most recent research efforts in hybrid networks concentrate on the integration of autonomous wireless networks (WLANs, WPANs, cellular) \cite{Mitseva2008}. These efforts put the focus on topology discovery and maintenance, cross-layer optimization and scalability \cite{aict}.

\section{Desiderata}
\label{sec:des}

The idea of identifying novel interaction paradigms which allow an optimal organization of resources both at global and local levels, thus overcoming the metacomputing limitations mentioned earlier, 
might appear rather ambitious and complicated. However, most of the technologies needed to achieve these target paradigms are already available; yet, they do not fully integrate with each other. 
These technologies include opportunistic, self-organizing networks and P2P interaction mechanisms. In addition, trust, security and privacy concerns are mandatory issues that need to be addressed.\\

Thus, the main objective to pursue is a novel distributed architecture that enables the development of ubiquitous applications, where all devices available to a user are dynamically and adaptively configured depending on: i) the devices themselves, ii) the environment in which they are deployed, and iii) the other users (and their characteristics) and the interactions they have. The architecture must thus provide configuration protocols for the intra Personal Area Networking (PAN), to automatically organize all devices belonging to a single human being seen as a DO. At the same time, it is necessary to identify algorithms and mechanisms for the simultaneous and adaptive use of different communication networks in an opportunistic fashion. In fact, the overall goal is to optimize the interactions across all DOs populating the ecosystem. Once the organism is able to make the best use of its devices within the digital ecosystem, it becomes necessary to develop protocols for the organization and interaction 
between different DOs. In other terms, mechanisms for the efficient aggregation and sharing are needed. 

\section{Design Issues}
\label{sec:archi}

In this section we introduce the principal design issues that need to be addressed in order to meet our desiderata. These issues are discussed below in isolation.


\subsection{Optimizing the Digital Organism}

Full interaction among all computing resources (and in general all hardware) belonging to the user is needed. This requires to optimize the use of networks that are available to the user's devices. For these interactions it is natural to adopt short-range communication technologies, such as Bluetooth, infrared, ZigBee, etc. This can be achieved through heuristic methods, but also (and perhaps most importantly) by means of decisional procedures based on several criteria. Key metrics are, of course, the computational capability of each device, the available bandwidth and the number of simultaneously interacting devices.\\

The goal is to model each DO as a computational environment and to find the best configuration for it. Specifically, based on the computational capacities of each device within the PAN, the battery levels, and the available network interfaces, devices must be configured so as to identify a primary computation entity, a primary gateway to send/receive data from the outside world, secondary network interfaces (e.g.~short range ones) to allow communications with neighbor organisms.\\

\begin{algorithm}[t]
\caption{\textsf{Configuration}}\label{alg:conf}
\begin{small}
\begin{algorithmic}[1]
\REQUIRE $myId$ device id
\REQUIRE $uId$ user id 
\REQUIRE $myProfile$ profile of $myId$
\FOR{each network interface $i$}
 \STATE{devs[$i$] = discoveryDeviceList($myId$,$uId$)}
 \STATE{send(devs[$i$], $myProfile$)}
 \STATE{profiles[$i$] = receiveFromAll()}
\ENDFOR
\STATE{}
\STATE{createInternalOverlay(devs, profiles)}
\STATE{coordinator = electCoordinator(profiles)}
\STATE{gateway = electGateway(profiles)}
\end{algorithmic}
\end{small}
\end{algorithm}

Algorithm \ref{alg:conf} reports the general scheme that devices, forming the DO, should execute during the configuration phase of the PAN. In essence, the idea is that devices must discover all other devices belonging to the same user $uId$. Since devices communicate through their available network interfaces, each device must concurrently perform such a discovery process. In fact, a given device, having two (or more) network interfaces, might act as a relay to interconnect other devices embodying different networking technologies. Hence, a first goal of this configuration phase is to create an intra-PAN communication overlay 
on heterogeneous networks.\\

An important aspect is that all nodes must distribute all their technical details among each other, in order to enable a proper configuration of the DO. Different alternatives exist to characterize profiles of devices, such as, for instance, CC/PP \cite{CCPP}. Such information is exploited to identify the coordinator, i.e.~the device that acts as the resource manager of the DO. To accomplish this task, all devices' profiles must be distributed among the whole device set internal to the DO, and some distributed algorithm must be executed to elect the coordinator. A similar approach must be employed to identify which device is to act as the primary gateway that manages communications with the outside world.

\subsection{Optimizing the Digital Organism's Interactions within the Digital Ecosystem}

The DOs must be provided with a set of protocols to interact with peer DOs within their home ecosystem. These protocols would allow a DO to opportunistically and dynamically adapt the interaction with its peer DOs, by selecting the best communication protocol among the available ones (e.g. Always Best Connected, ABC) \cite{wirelessdays}. The identification of the best available network may be based on multiple criteria such as: i) bandwidth, ii) connection cost, iii) battery consumption, iv) probability of maintaining the connection active while moving, in order to minimize the hand-offs. Any of these criteria, alone or together with the others, can be used for assessing the best available network at any time. A much more challenging target is to provide techniques and methods for the simultaneous and opportunistic use of all available multiple networks \cite{aict}. 

By redirecting traffic across multiple network interfaces one may suitably balance each network's load. This approach has the fundamental advantage of permitting communication even when one of the various operating networks suddenly falls. To achieve all this, it is important to employ suitable communication solutions which, for example, make use of proxies and cross-layer mechanisms in order to firstly distributing the information flow over multiple communication channels and next re-aggregating the whole flow together again \cite{aict}. This seems the only way for making the use of multiple network interfaces transparent at the application level. In fact, the above process should be fully transparent to the final users (with the exception of security and privacy issues, which may require an explicit user intervention in order to be dealt with).\\

\begin{figure}[t]
\centering
 \includegraphics[angle=-90,width=\linewidth]{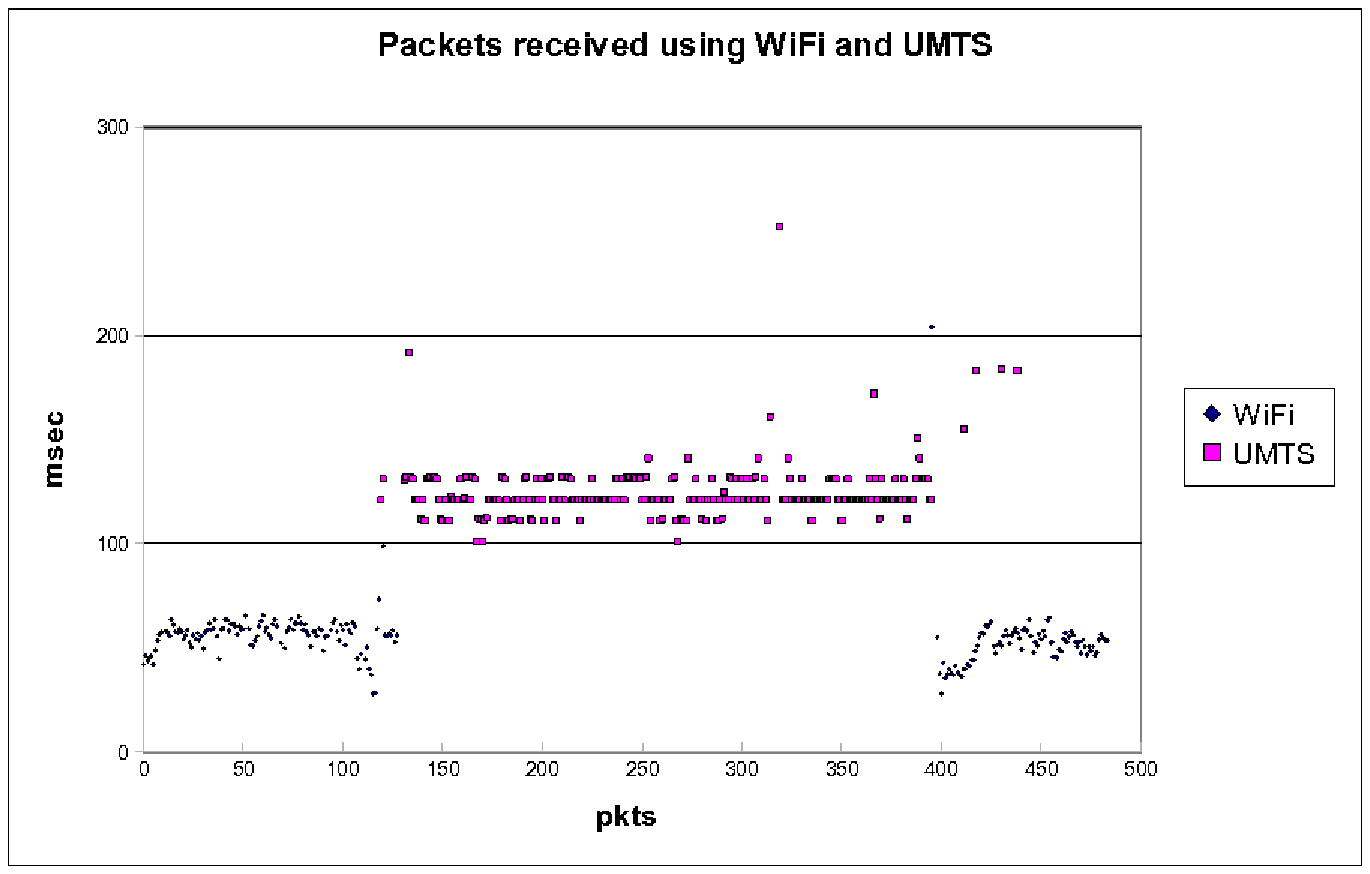}
\caption{Seamless connectivity obtained using multiple communication technologies}
\label{fig:cross_layer}
\end{figure}

Figure \ref{fig:cross_layer} illustrates how effective the use of multiple networks can be in order to provide seamless communications. The chart refers to the use of a (really implemented) cross-layer architecture that uses a proxy-based system to offer continuity in the communication of a given mobile node with a remote proxy \cite{aict}. The mobile node chart was equipped with two different network interfaces, i.e, a Wi-Fi card and a UMTS card. During the path performed by the mobile node, a temporary unavailability of the Wi-Fi network was experienced. Nevertheless, the mobile node was able to automatically switch to the UMTS network, without interrupting the end-to-end communication with the remote node. This was possible using a proxy which hides the handover from a network to another (and hence, also the change of IP address for the mobile node). In the figure, the experienced latencies for packets transmissions are reported; needless to say, those related to packets delivered through UMTS are higher 
than those sent via Wi-Fi. Nevertheless, a continuous communication was guaranteed to the application \cite{aict}. This represents just a (real) example that demonstrates how viable solutions can be employed to optimize the interactions of the DOs, if it is able to automatically employ all the resources which are available (in this case, two different types of networks), transparently to its human user.

\subsection{Optimizing the Digital Ecosystem}
\label{sec:deco}

The next step is to organize the DOs as a network of interacting nodes. Many approaches to achieve this goal are possible such as overlay or mesh networks, or in certain settings mechanisms based on not fully decentralized approaches \cite{androu}. The main aim is at modeling nodes' interactions while taking full advantage of the typical paradigms in terms of complex networks and adaptive systems. In this perspective, the theory of complex networks might help to identify the best topology to organize the users' interactions. For instance, in certain scenarios, the use of random networks with (almost) uniform nodes' degrees could be employed, while in other contexts ``small worlds'' and scale-invariant networks could be better choices \cite{simutools}.\\

Moreover, nodes need not only be able to organize themselves at random, but also and most importantly according to local strategies based on nodes' targets, resources and services (both demanded and offered). Other parameters for deciding whether to create a link or not include the human identities behind the nodes. More specifically, given the privacy concerns raised by some data to be routed through the network, it is natural that connections will be preferably created among trustworthy DOs.\\

This task is easy to handle if the network configuration is left to the user. Conversely, if the goal is to obtain a (semi)~automatic configuration mechanism, then it becomes crucial (and hard) to implement an approach based on the grouping of nodes (e.g. using metrics similar to those employed in social applications). In such a case, network configuration becomes a non-negligible problem, in that one has to simultaneously take into account: i) nodes' topology in the environment, ii) the type of network to be created, iii) the type of services that nodes are actually seeking/offering, iv) node heterogeneity in terms of communication and computation capabilities, v) the clustering/proximity among nodes that are close to each other in terms of social features.\\

Dissemination strategies must be adopted to distribute messages. In a wide and heterogeneous network such as that under consideration, it is important to obtain smart algorithms that allow to broadcast messages among the network nodes. Such a functionality would allow to implement techniques for the discovery of nodes, resources, paths to transmit messages towards a destination. It is important to notice that, due to the high dynamic of the network, such schemes might be based on unstructured P2P solutions \cite{gridpeer,guclu}. Gossip dissemination strategies are thus extremely important in this contexts \cite{HaridasanvanRenesse08}.\\

\begin{figure*}[t]
\centering
 \includegraphics[angle=-90,width=0.48\linewidth]{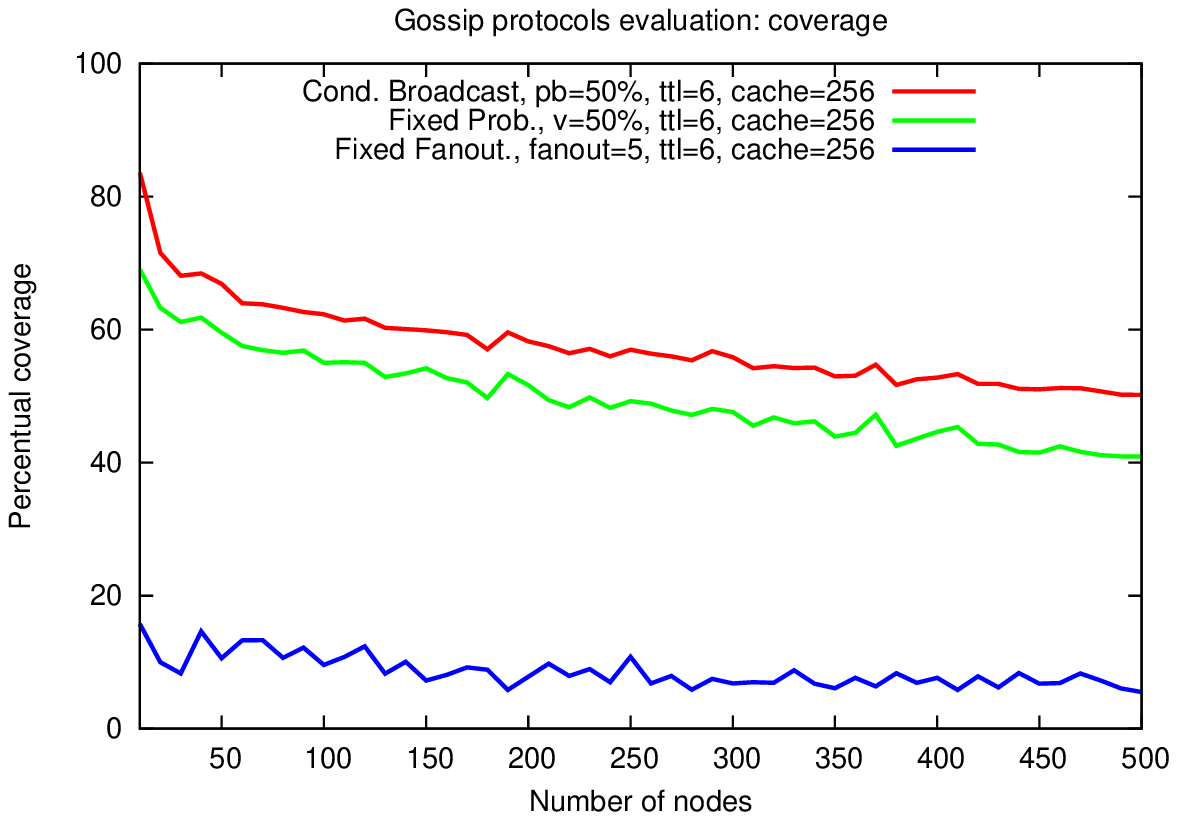}
 \includegraphics[angle=-90,width=0.48\linewidth]{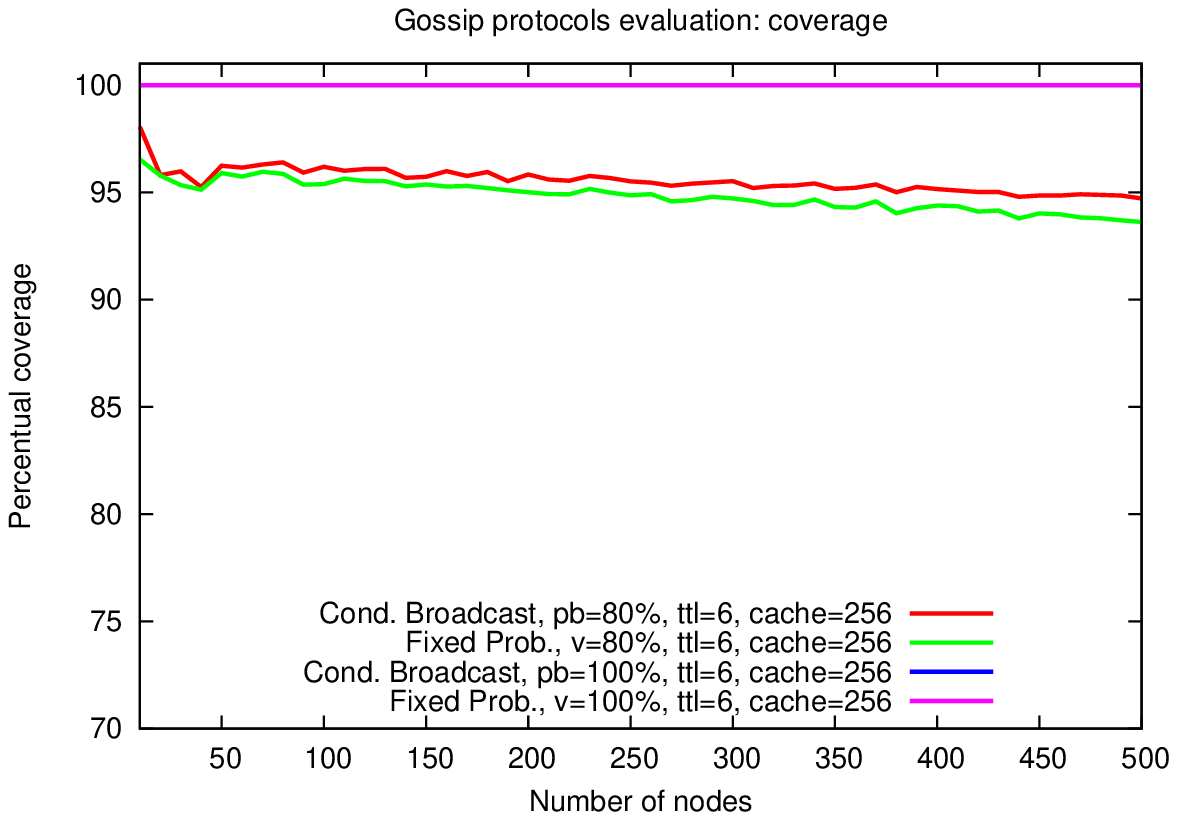}
 \includegraphics[angle=-90,width=0.48\linewidth]{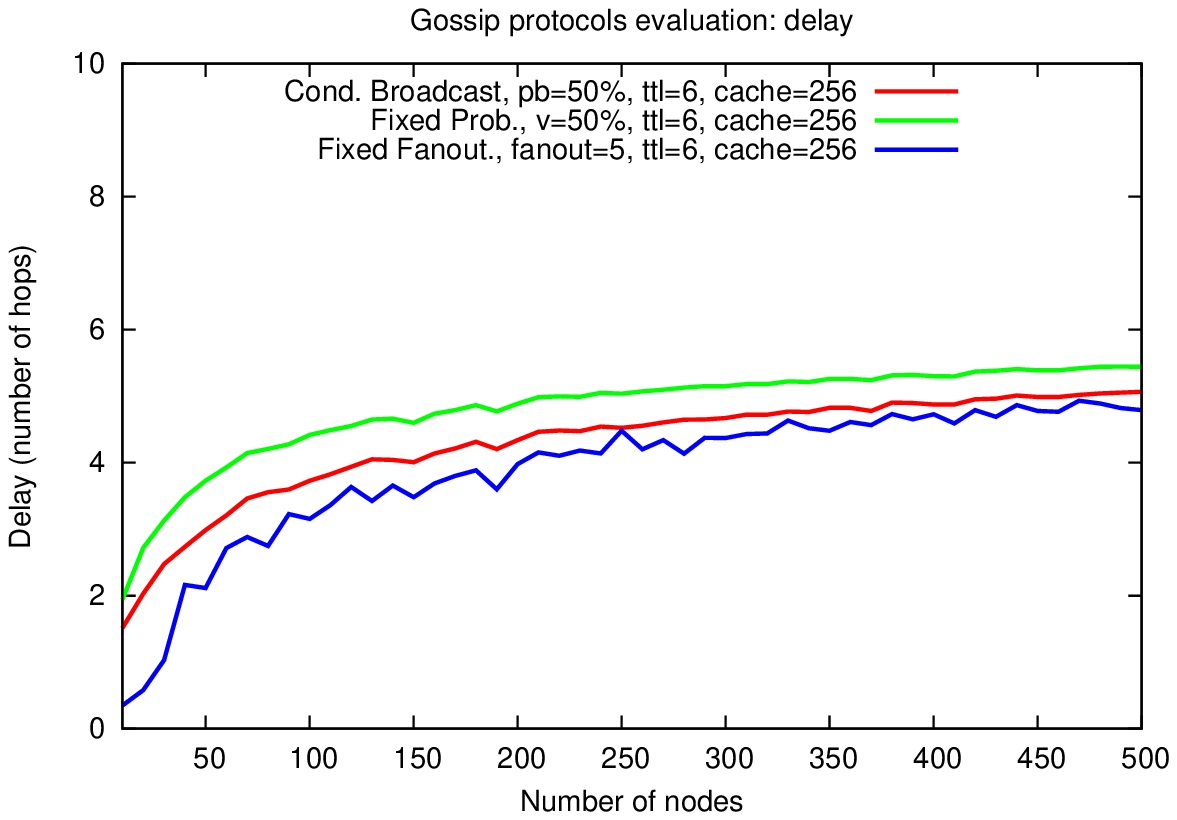}
 \includegraphics[angle=-90,width=0.48\linewidth]{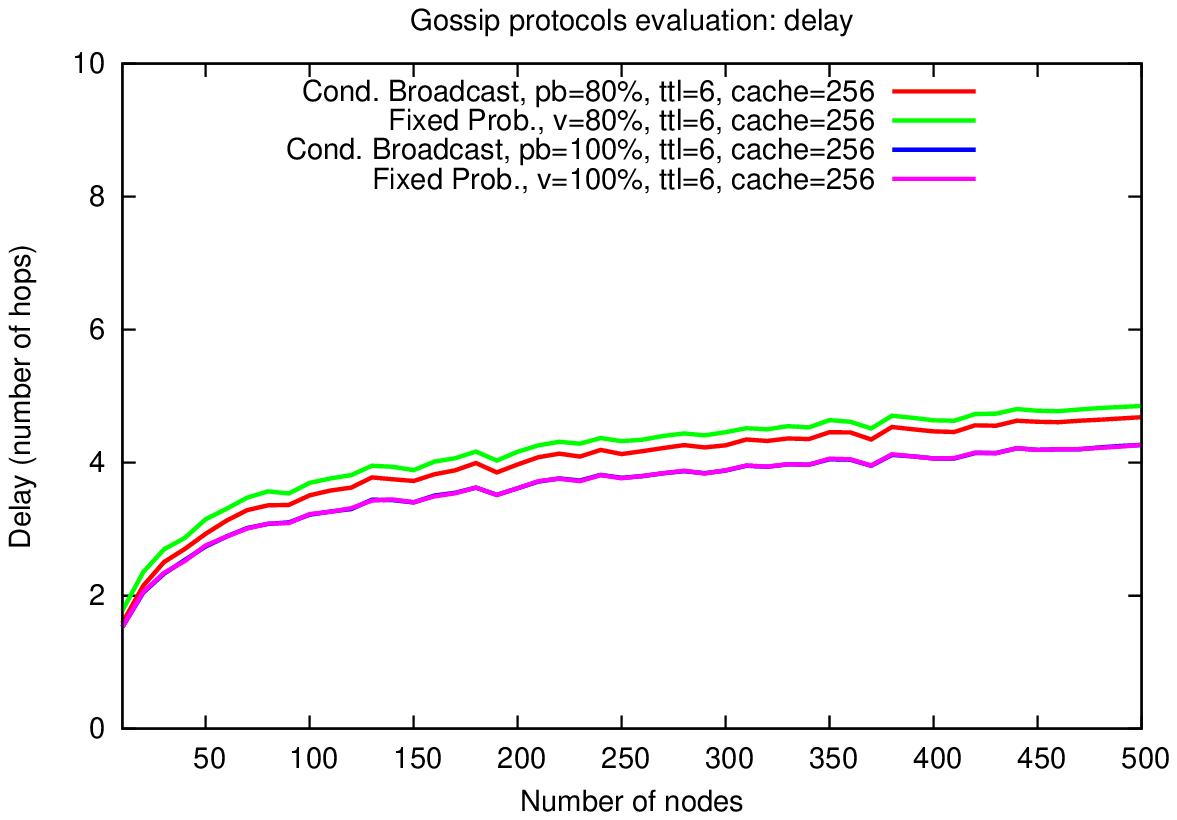}
\caption{Performances of different gossip schemes. Coverages and number of hops for different probabilities of dissemination}
\label{fig:gossip}
\end{figure*}

Figure \ref{fig:gossip} demonstrates that, when adequately tuned, gossip algorithms allow to disseminate messages in a network. Three examples of gossip protocols are considered: i) a message arriving at a node is forwarded to all the node neighbors with a given probability (Conditional Broadcast); ii) a message arriving at a node is forwarded to each neighbor with a given probability (Fixed Probability); iii) a message arriving at a node is forwarded to a fixed number of (randomly selected) neighbors (Fixed Fanout) \cite{disio}. The charts report the average coverage of message dissemination (i.e.~the percentage of nodes that receive a specific broadcast) and the average delay (i.e.~number of hops), depending on the dissemination probability, the time to live (ttl) exploited for disseminated messages, and the size of cache set at nodes.

\subsection{Optimizing the Computation}

Once created the infrastructure for communication and interaction, one may handle all computationally demanding queries in an adaptive manner, thereby implementing a true cloud computing system over mobile ad-hoc technologies \cite{Babaoglu:2006}. Users may require computations to be performed on remote and/or unknown hosts, trusted servers, or hosts belonging to the same user domain. To this end, one firstly needs to identify/discover those hosts which are able to handle the submitted requests, and next identify the best way to submit it. In this framework, necessary properties are privacy, accountability and non repudiability. In non-mobile scenarios, these issues have been studied at length; within grid and cloud computing, ad-hoc processing and P2P computing systems, there are many well established useful results \cite{ma,Reyes:2010}. 
In contrast, in the context of mobile computing, these issues lack universally accepted solutions and are still under investigation.

\section{Conclusions}
\label{sec:conc}

In this paper, we have discussed a methodology to optimize the interactions of mobile users (in the paper referred as digital organisms) into dynamic and heterogeneous environments (termed digital ecosystem). The idea is to optimize the use and interaction of the devices available to each digital organism, through dynamic and adaptive configuration strategies (optimization of the PAN). Specifically, the interactions within a DO and among DOs can be optimized using both the available communication infrastructures and P2P ad-hoc interactions. 

Our next step will be to complete the implementation of our approach.
We expect to be able to report on our results in the near future.



\small{
\bibliographystyle{abbrv}
\bibliography{biblio}  
}


\end{document}